# A kinetic perspective of charge transfer reactions: the downfall of hard/soft acid/base interactions

Ramón Alain Miranda-Quintana

Department of Chemistry & Quantum Theory Project; University of Florida; Gainesville, FL 32611, USA

Email: quintana@chem.ufl.edu

We show how to incorporate the possibility of kinetic control in the conceptual Density Functional Theory formalism. This allow us to prove that the hard/soft acid/base principle will likely fail when the reactions are not thermodynamically-driven.

Chemical intuition still plays a paramount role in our understanding of reactivity. Despite the widespread adoption of robust computational techniques allowing us to scan a vast number of reactions in search for their potential products, in the end, chemists are not satisfied until we understand the factors driving the formation of those compounds. In the words of Coulson, we seek "primitive patterns of understanding". Initially confined to rationalizing multiple experimental results, the quest for these reactivity principles has steadily shifted towards a more theoretical approach. Molecular orbitals have been a very popular tool in this regard, with many chemists effectively thinking in terms of single-particle energy levels, frontier orbitals, etc.[1-7] However, approaches based on Density Functional Theory[8,9] (DFT) are arguably more powerful, because by being based on an observable (e.g., the electron density) it is easier to connect them to different fundamental physical principles. This area of research, commonly termed conceptual DFT (C-DFT)[10-19] has had great success explaining well-established reactivity rules, like Sanderson's electronegativity equalization,[20,21] or even helping establish totally new ones, like the maximum hardness,[22-25] minimum electrophilicity,[26-33] or the "$|\Delta\mu|$ big is good"[9,34-39] principles. However, arguably the crown jewel of C-DFT is the first-principles validation of Pearson's hard/soft acid/base (HSAB) rule.[40-42] With several proofs of this basic principle,[43-45] C-DFT seems to have firmly established that, indeed, hard acids tend to prefer to bind to hard bases, while soft bases tend to prefer soft acids. But, there are plenty of instances when this is not the observed behavior. Within the C-DFT community the most common justifications are that the "elementary" proofs of HSAB introduce many approximations (e.g., non-realistic molecular interaction models, simplistic charge transfer models without high-order terms) that leave plenty of room open for HSAB to be broken in real conditions. The counterargument to this, on the other hand, is that in recent times there have been several slightly more elaborated proofs of HSAB that include these effects.[46-49] In this contribution we take a look at this contradiction: why HSAB fails in so many cases, even when C-DFT says it should be way more prevalent? The key insight explored below was provided by Mayr and others,[50-52] when they noted that the standard HSAB treatment seems to be agnostic of the notions of thermodynamic and kinetic control.

All basic attacks to the HSAB problem demand two things: 1) a way to model energy changes along a reaction; 2) a definition of chemical hardness. Parr and Pearson solved both of these issues at the same time, relating the latter to the former via their venerable quadratic model:[53]

$$\Delta E = \mu \Delta N + \frac{1}{2}\eta \Delta N^2 \qquad (1)$$

Here, $\mu = \left(\frac{\partial E}{\partial N}\right)_{v(\mathbf{r})}$ is the electronic chemical potential[54] (which represents how the energy of a system will change with respect to changes in the number of electrons, $N$, so it can be identified with the negative of the electronegativity), and $\eta = \left(\frac{\partial^2 E}{\partial N^2}\right)_{v(\mathbf{r})}$ is the chemical hardness[53] (a measure of the resistance encountered to accept/donate electrons). This expression can then be used to estimate the change in energy along a prototypical acid-base (e.g., charge transfer) reaction $A + B = AB$:

$$\Delta E = -\frac{1}{2}\frac{(\mu_B - \mu_A)^2}{\eta_A + \eta_B} \qquad (2)$$

Moreover (and without losing any generality), if we consider that $B$ is the base and $A$ is the acid, we can even calculate the amount of charge transferred:

$$\Delta N = \frac{\mu_B - \mu_A}{\eta_A + \eta_B} \qquad (3)$$

(Notice that, since we can use the chemical potential as a measure of the tendency to donate electrons: $\mu_B > \mu_A$.)[45,55]

These expressions are all that is typically needed to prove the HSAB principle. This is usually done by considering a double-exchange reaction:

$$A_hB_s + A_sB_h \leftrightarrow A_sB_s + A_hB_h \tag{4}$$

where the sub-indices denote if the species are hard (*h*) or soft (*s*).

Then, we need to make some assumptions:

I: $\mu_{A_s} = \mu_{A_h} = \mu_A$; $\mu_{B_s} = \mu_{B_h} = \mu_B$

II: $\eta_{A_s} = \eta_{B_s} = \eta_s$; $\eta_{A_h} = \eta_{B_h} = \eta_h$

With this, the reaction energy of Eq. (4) reduces to:

$$\Delta E = -\frac{1}{2}(\mu_A - \mu_B)^2 \left( \frac{1}{2\eta_h} + \frac{1}{2\eta_s} - \frac{2}{\eta_h + \eta_s} \right) < 0 \tag{5}$$

As indicated by the inequality in Eq. (5), the formation of the soft-soft and hard-hard adducts will be more favored, thus validating HSAB. However, the reaction energy will be the defining criterion determining the reactivity if the reaction is controlled thermodynamically. This begs the question of how to analyze the double exchange reaction if it proceeds under kinetic control.

The first step to assess the relevance of kinetic control in charge transfer reactions is to have a model to estimate the activation energy of these processes. The key insight here is that we can separate the reaction energy given in Eq. (2) in terms of the contribution for each reactant:[56-58]

$$\Delta E = \Delta E_{A(B)} + \Delta E_{B(A)} \tag{6}$$

$$\Delta E_{A(B)} = \Delta N \left( \mu_A + \frac{1}{2}\eta_A \Delta N \right) < 0 \tag{7}$$

$$\Delta E_{B(A)} = \Delta N \left( -\mu_B + \frac{1}{2}\eta_B \Delta N \right) > 0 \tag{8}$$

Notice how the energetic contribution of the acid is always going to favor the formation of the product, while the contribution of the base goes "uphill", opposing this process. Hence, it is natural to identify the latter with the activation energy of the reaction. This notion has previously been successfully used to rationalize several families of reactions, even providing quantitative agreement with experimental and high-level theoretical calculations. In the following, we will then denote $E_a \equiv \Delta E_{B(A)}$.

Our next step is then to study the behavior of this activation energy for the two distinct possibilities: the formation of the pro-HSAB or anti-HSAB products. For simplicity, we will consider that the reactions occur in a concerted way, so we can effectively add the contributions of the different activation energies to each overall reaction. Moreover, we will be making the same assumptions regarding the chemical potential and hardnesses of the involved species as detailed above. These are arguably the most pro-HSAB conditions possible, so if anything by choosing them we are considering the cases in which the HSAB principle should be expected to hold.

Case 1: pro-HSAB reaction

$$(A_h + B_h) + (A_s + B_s) \rightarrow A_h B_h + A_s B_s \qquad (9)$$

$$\begin{aligned}
E_a^{(\text{pro})} &= E_a^{(A_h B_h)} + E_a^{(A_s B_s)} \\
&= \frac{\mu_B - \mu_A}{2\eta_h}\left\{-\mu_B + \frac{\eta_h}{2}\frac{(\mu_B - \mu_A)}{2\eta_h}\right\} + \frac{\mu_B - \mu_A}{2\eta_s}\left\{-\mu_B + \frac{\eta_s}{2}\frac{(\mu_B - \mu_A)}{2\eta_s}\right\} \\
&= \frac{\mu_B - \mu_A}{2}\left\{-\mu_B + \frac{(\mu_B - \mu_A)}{4}\right\}\left(\frac{1}{\eta_h} + \frac{1}{\eta_s}\right)
\end{aligned} \qquad (10)$$

Case 2: anti-HSAB reaction

$$(A_h + B_h) + (A_s + B_s) \rightarrow A_h B_s + A_s B_h \qquad (11)$$

$$E_a^{(\text{anti})} = E_a^{(A_hB_s)} + E_a^{(A_sB_h)}$$

$$= \frac{\mu_B - \mu_A}{\eta_s + \eta_h}\left\{-\mu_B + \frac{\eta_h}{2}\frac{(\mu_B - \mu_A)}{\eta_s + \eta_h}\right\} + \frac{\mu_B - \mu_A}{\eta_s + \eta_h}\left\{-\mu_B + \frac{\eta_s}{2}\frac{(\mu_B - \mu_A)}{\eta_s + \eta_h}\right\} \quad (12)$$

$$= \frac{\mu_B - \mu_A}{2}\left\{-\mu_B + \frac{(\mu_B - \mu_A)}{4}\right\}\left(\frac{1}{\eta_s + \eta_h}\right)$$

Then:

$$\frac{E_a^{(\text{pro})}}{E_a^{(\text{anti})}} = \frac{\left(\dfrac{1}{\eta_h} + \dfrac{1}{\eta_s}\right)}{\left(\dfrac{1}{\eta_s + \eta_h}\right)} = 2 + \frac{\eta_h^2 + \eta_s^2}{\eta_h \eta_s} \quad (13)$$

From here (and using the fact that the hardnesses are always positive, due to the convexity of the energy),[59-62] we trivially get that:

$$\frac{E_a^{(\text{pro})}}{E_a^{(\text{anti})}} = 2 + \frac{\eta_h^2 + \eta_s^2}{\eta_h \eta_s} > 1$$
$$E_a^{(\text{pro})} > E_a^{(\text{anti})} \quad (14)$$

In other words, if the acid base double-exchange reaction is kinetically controlled, the main products are going to be hard-soft adducts, in direct contradiction with the HSAB rule. Once again, it is important to remember that this result was obtained under conditions that actually were biased towards the validity of HSAB, hence the prevalence of the mix-hardness products is all the more remarkable.

The obvious main conclusion of this manuscript was already anticipated in its title: the HSAB principle should not be expected to hold in kinetically controlled reactions. The common dictum that "all other things being equal, hard acids prefer binding to hard bases when soft acids are binding to soft bases" should be appended by "under conditions of thermodynamic control". While at first this might seem like a negative result, we actually think it presents an encouraging extension to the traditional C-DFT formalism. A C-DFT "proof" that HSAB also follows from kinetic control would have been in obvious contradiction with ample experimental evidence, thus demanding a deeper look into potential inadequacies in the theoretical framework. What we have shown here, on the other hand, is that the C-DFT formalism is capable of describing conditions in which even the most respected of its principles can fail. This is an even more reassuring result, speaking of the robustness of the physical and mathematical structure of C-DFT. Of course, a full analysis of the HSAB principle cannot be just confined to a simple parabolic charge transfer dependency. While the insights provided here point out to clear reasons to expect the failure of HSAB in several cases, it is certainly desirable to extend these results to more realistic conditions. Likewise, we hope that this work brings attention to the importance of clearly separating thermodynamically and kinetically controlled reactions in the C-DFT framework, with special emphasis on revisiting the validity of other reactivity principles in the latter regime.

## ACKNOWLEDGMENTS

We thank the University of Florida for support in the form of a startup grant.